\documentclass[useAMS,usenatbib]{mn2e}

\pdfoutput=1

\usepackage{graphicx}
\usepackage{subfig}
\usepackage{amssymb}

\usepackage{aas_macros}
\usepackage{times}

\newcommand{\Hii}{{\sc H$\,$ii} }

\title[Proto-galaxies reionising the Universe]{The First Billion Years project - IV: Proto-galaxies reionising the Universe}
\author[J.-P. Paardekooper et al.]{Jan-Pieter Paardekooper$^{1}$\thanks{E-mail:
jppaarde@mpe.mpg.de}, Sadegh Khochfar$^{1}$ and Claudio Dalla Vecchia$^{1}$ \\
$^{1}$Max Planck Institute for extraterrestrial Physics, PO Box 1312, Giessenbachstr., 85741 Garching, Germany}
\begin{document}

\date{Accepted ***. Received ***; in original form ***}

\pagerange{\pageref{firstpage}--\pageref{lastpage}} \pubyear{2012}

\maketitle

\label{firstpage}

\begin{abstract}
%Context
The contribution of stars in galaxies to cosmic reionisation depends on the star formation history in the Universe, the abundance of galaxies during reionisation, the escape fraction of ionising photons and the clumping factor of the inter-galactic medium (IGM).
%Aims
%Methods
We compute the star formation rate and clumping factor during reionisation in a cosmological volume using a high-resolution hydrodynamical simulation. We post-process the output with detailed radiative transfer simulations to compute the escape fraction of ionising photons. Together, this gives us the opportunity to assess the contribution of galaxies to reionisation self-consistently.
%Results
The strong mass and redshift dependence of the escape fraction indicates that reionisation occurred between $z=15$ and $z=10$ and was mainly driven by proto-galaxies forming in dark-matter haloes with masses between $10^7 \, \mathrm{M}_{\odot}$ and $10^8 \, \mathrm{M}_{\odot}$. More massive galaxies that are rare at these redshifts and have significantly lower escape fractions contribute less photons to the reionisation process than the more-abundant low-mass galaxies. Star formation in the low-mass haloes is suppressed by radiative feedback from reionisation, therefore these proto-galaxies only contribute when the part of the Universe they live in is still neutral. After $z\sim10$, massive galaxies become more abundant and provide most of the ionising photons. In addition, we find that Population (Pop) III stars are too short-lived and not frequent enough to have a major contribution to reionisation. 
%Conclusions
Although the stellar component of the proto-galaxies that produce the bulk of ionising photons during reionisation is too faint to be detected by the {\it James Webb Space Telescope} (JWST), these sources are brightest in the $\mathrm{H}\alpha$ and Ly-$\alpha$ recombination lines, which will likely be detected by JWST in deep surveys.
\end{abstract}


\section{Introduction}

One of the major challenges in modern cosmology is identifying the nature of the sources responsible for reionising the Universe. The Gunn-Peterson trough in the spectra of high-redshift quasars indicates that the IGM was highly ionised at $z<6$ \citep{Fan:2006he}. The integrated Thomson (electron) scattering optical depth to the surface of last scattering, $\tau_{\mathrm{e}}$, suggests that reionisation was well underway by $z \approx 10.5$ \citep{2011ApJS..192...18K}. It remains uncertain which sources transformed the IGM into its highly ionised state. The most likely candidates are stars in galaxies \citep{2008ApJ...682L...9F}. 

The contribution of galaxies to reionisation depends critically on the star formation history in the Universe, the abundance of reionising sources, the clumping factor of the IGM and the fraction of ionising photons that escapes into the IGM, the so-called escape fraction. High-redshift surveys that are probing the star formation rate of galaxies at $6 \lesssim z \lesssim 8$ show that an escape fraction of more than $\sim 30\%$ is needed for the observed galaxy population to produce enough photons to keep the Universe ionised \citep{2010Natur.468...49R,2012ApJ...752L...5B}. Although there is some evidence for a redshift evolution of the escape fraction, such high escape fractions are not observed in the local Universe \citep{BlandHawthorn:1999jm,Deharveng:2001gr,Heckman:2001fn} and at higher redshifts \citep{Iwata:2009dy,2011ApJ...736...18N,2012ApJ...751...70V}. All the massive galaxies targeted by observations have average escape fractions less than 20\%, although samples are small. With this escape fraction, the ionising emissivity of the observed galaxy population is insufficient to maintain reionisation. However, the bulk of the star formation rate at these high redshifts likely arises in galaxies below the detection limit of current observational facilities \citep{2012ApJ...754...46T}. If the escape fraction evolves with redshift, these galaxies could provide the majority of photons for reionisation \citep{Trenti:2010fe,2012MNRAS.423..862K}. 

Numerical studies \citep{Gnedin:2008ib,Wise:2009fn,Razoumov:2010bh,Yajima:2010fb} find escape fractions between $\sim 0 - 1$, with possible redshift or mass dependence. In part the large differences between studies may be caused by numerical issues, because the radiative transfer simulations are computationally challenging, making approximations necessary. However, most studies targeted only a few objects and different studies focussed on different mass galaxies at different redshifts, making comparison difficult. In simulations of idealised, isolated galaxies \citet{Paardekooper:2011cz} found the main constraint on the escape fraction to be the dense gas in the star-forming regions, which provides an explanation for the large spread in escape fractions reported in previous studies. In addition, they found that the escape fraction can vary over several orders of magnitude over the lifetime of the galaxy, making it necessary to determine the contribution of galaxies to reionisation not only sampling the mass function, but also over a wide range of redshifts.

In this letter we present results on the escape fraction and ionising emissivity from a large statistical sample of proto-galaxies in a high-resolution cosmological, hydrodynamical simulation. In combination with the derived clumping factor of the IGM in the simulated volume we assess self-consistently the contribution of stars in galaxies to cosmic reionisation.

\section{Method}

In the standard cold dark matter paradigm, at the relevant redshifts for reionisation most ionising radiation is produced by stars in proto-galaxies forming in dark matter haloes of $<  10^9 \, \mathrm{M}_{\odot}$ \citep{Barkana:2001tz,Choudhury:2008dm}. We compute the ionising photon production and escape fraction in proto-galaxies in this mass range using the {\it First Billion Years} (FiBY) simulation suite (Khochfar et al. in prep.; Dalla Vecchia et al. in prep.). The simulation we use contains $2 \times 684^3$ dark matter and gas particles in a comoving volume of 4 Mpc on the side, with a gas-particle mass of 1250 $M_{\odot}$. At redshift 6, the simulation reproduces the observed mass function of galaxies and star formation rates (Khochfar et al. in prep.). 

For the FiBY simulation we use a modified version of the {\it OWLS} code \citep{Schaye:2010jl}. Star formation follows a pressure law \citep{2008MNRAS.383.1210S}, where we assume population (Pop) III stars form at metallicities $Z < 10^{-4} \, \mathrm{Z}_{\odot}$ (with $\mathrm{Z}_{\odot} = 0.02$) and Pop II stars at higher metallicities. Supernova feedback is modelled by injecting thermal energy that is efficiently converted into kinetic energy without the need to turn off radiative cooling temporarily \citep{2012MNRAS.426..140D}. Feedback from reionisation is modelled as a uniform UV-background following \citet{2001cghr.confE..64H} by switching from collisional to photo-ionisation equilibrium cooling tables. Gas above a density threshold of $n_{\mathrm{shield}} = 0.01 \, \mathrm{cm}^{-3}$ is modelled to shield against ionising radiation \citep{2011AAS...21734501N}. We assume that reionisation takes place within the bounds set by WMAP \citep{2011ApJS..192...18K}, starting around redshift $12$ and ending around redshift $9$. This is consistent with the computations of the ionising emissivity from proto-galaxies  that we present in this letter. We will show that between redshift $12$ and $9$ the proto-galaxies in the simulation produce enough ionising photons to reionise the computational volume.

We have extracted all haloes from this simulation that contain at least 1 star, 1000 dark matter particles and 100 gas particles for post-processing with radiative transfer simulations. This results in more than 11000 haloes between $z=20$ and $z=6$. We determine the escape fraction with an updated version of the {\sc SimpleX} radiative transfer code \citep{Paardekooper:2010iu}, computing the absorption of the ionising radiation by both hydrogen and helium atoms in 10 frequency bins until the photons reach the virial radius of the halo. We will discuss these simulations in more detail in a forthcoming paper (Paardekooper et al. in prep). The luminosity and spectra of the star particles are computed from stellar synthesis models for both Pop III \citep{Raiter:2010hs} and Pop II \citep{2003MNRAS.344.1000B} stars. Combined with the escape fraction, we obtain the number of ionising photons that every proto-galaxy in our simulation contributes to cosmic reionisation.

To first order the reionisation process can be modelled by equating the number of photons produced per baryon to the number of recombinations in the ionised IGM \citep[e.g.][]{Madau:1999kl}. The volume fraction of ionised hydrogen, $Q_\mathrm{H \, II}$ is then given by
\begin{equation} %volume filling fraction of ionised hydrogen
  \frac{\mbox{d} Q_\mathrm{H \, II} }{\mbox{d}t} = \frac{ \dot{N}_{\mathrm{ion}}}{\bar{n}_{\mathrm{H,}0}} - Q_\mathrm{H \, II} \, C \, \bar{n}_{\mathrm{H,}0} \, \alpha(T) \, (1+z)^3,
\end{equation}  
where $\dot{N}_{\mathrm{ion}}$ is the total number of ionising photons available for reionisation per second per comoving Mpc, $\bar{n}_{\mathrm{H,}0} = 1.90641 \times 10^{-7} \, \mathrm{cm}^{-3}$  is the current mean number density of hydrogen in the IGM, $\alpha(T)$ is the recombination coefficient of hydrogen, which is a function of the IGM temperature $T$, $C \equiv \langle n_{\mathrm{H}}^2 \rangle / \langle n_{\mathrm{H}} \rangle^2$ is the clumping factor of the gas in the IGM and $z$ is the redshift. We assume that the ionised gas in the IGM has a temperature of 20,000 K, while we compute the clumping factor of the IGM gas from the simulation, using only gas with overdensity $\Delta < 100$, thus excluding gas inside dark matter haloes (because recombinations in that gas are already taken into account in the escape fraction calculations). We find a redshift-dependent clumping factor between 1.5 and 6.5, consistent with previous studies \citep{Pawlik:2009id,2012ApJ...747..100S}. We compute the Thomson optical depth, which is the quantity measured by the WMAP satellite, by integrating $Q_\mathrm{H \, II}$ over all redshifts:
\begin{equation} %Thomson optical depth
  \tau_{\mathrm{e}} = \int_0^{z_{\mathrm{rec}}} \mbox{d}z \left| \frac{ \mbox{d} t}{ \mbox{d} z} \right| c \, Q_\mathrm{H \, II}(z) \, \bar{n}_{\mathrm{H,}0} \, (1+z)^3 \, \sigma_{\mathrm{T}},
\end{equation}  
with $\sigma_{\mathrm{T}}$ the cross section for Thomson scattering and $c$ the speed of light.

\section{The contribution of proto-galaxies to cosmic reionisation}

\begin{figure} %the escape fraction as function of redshift
  \includegraphics[width=\columnwidth]{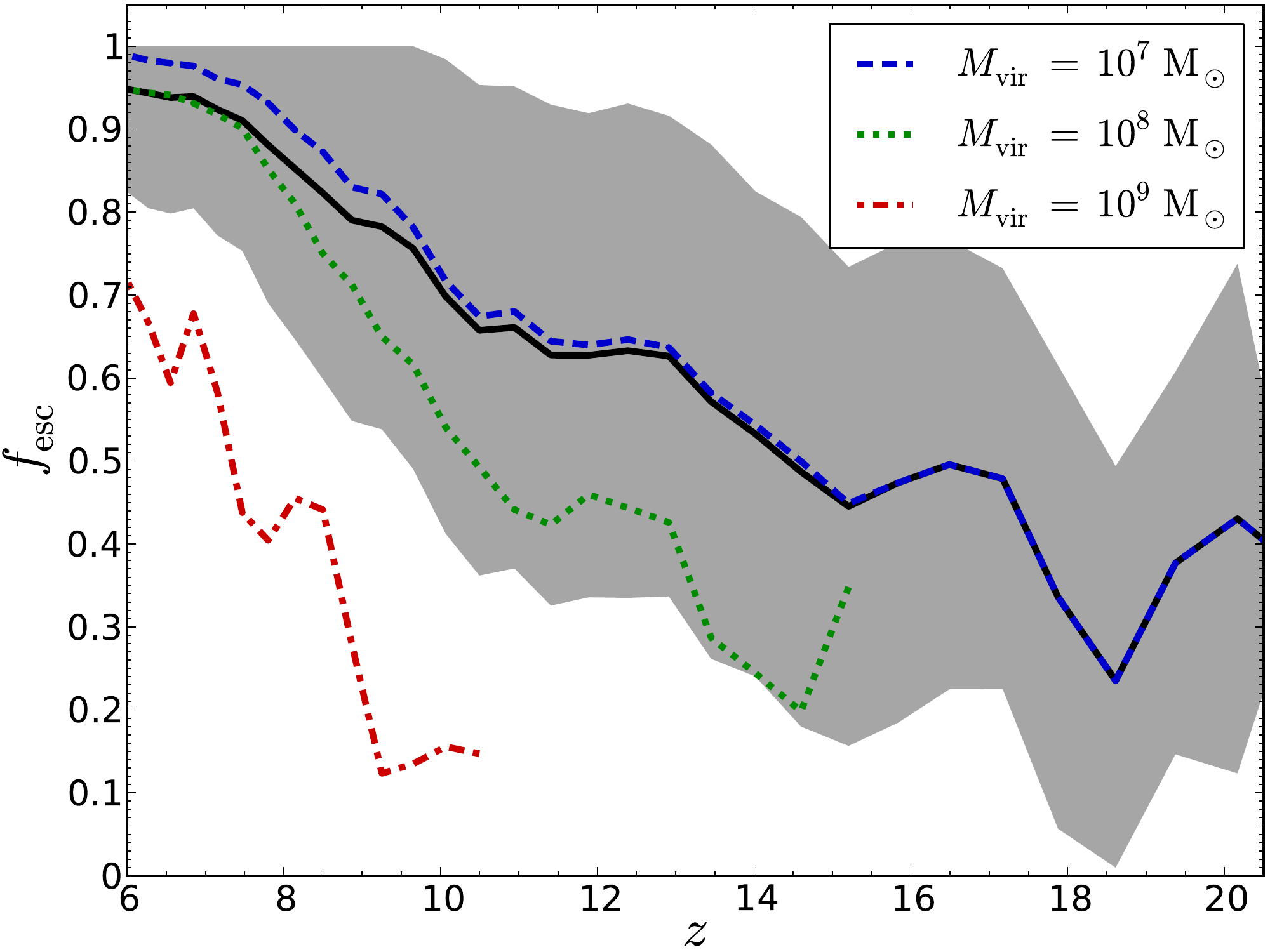}
  \caption{The escape fraction of ionising photons as function of redshift. The black solid line represents the escape fraction averaged over all haloes, while the blue dashed, green dotted and red dash-dot lines denote the escape fraction averaged over proto-galaxies in haloes with virial masses of $10^7 \, \mathrm{M}_{\odot}$, $10^8 \, \mathrm{M}_{\odot}$ and $10^9 \, \mathrm{M}_{\odot}$, respectively. The grey area represents the standard deviation of the mean. 
}
  \label{fig_fEsc}
\end{figure}

In Fig. \ref{fig_fEsc} we show the average escape fraction as function of redshift for proto-galaxies in haloes of different masses. The average escape fraction rises with time, but proto-galaxies inside a certain mass halo at the same redshift may have very different escape fractions, for example due to a different formation history or environment, which causes the large standard deviation in the mean. Due to the efficiency of stellar feedback in clearing away the gas from the dense sites of star formation, the escape fraction in $10^7 \, \mathrm{M}_{\odot}$ haloes is higher than in the $10^8 \, \mathrm{M}_{\odot}$ haloes. Proto-galaxies in haloes with masses above $10^9 \, \mathrm{M}_{\odot}$ have, at all redshifts, significantly lower escape fractions than their counterparts in lower mass haloes, due to their larger and denser gas content.

\begin{figure} %the volume filling fraction of ionised hydrogen and optical depth for Thomson scattering
  \includegraphics[width=\columnwidth]{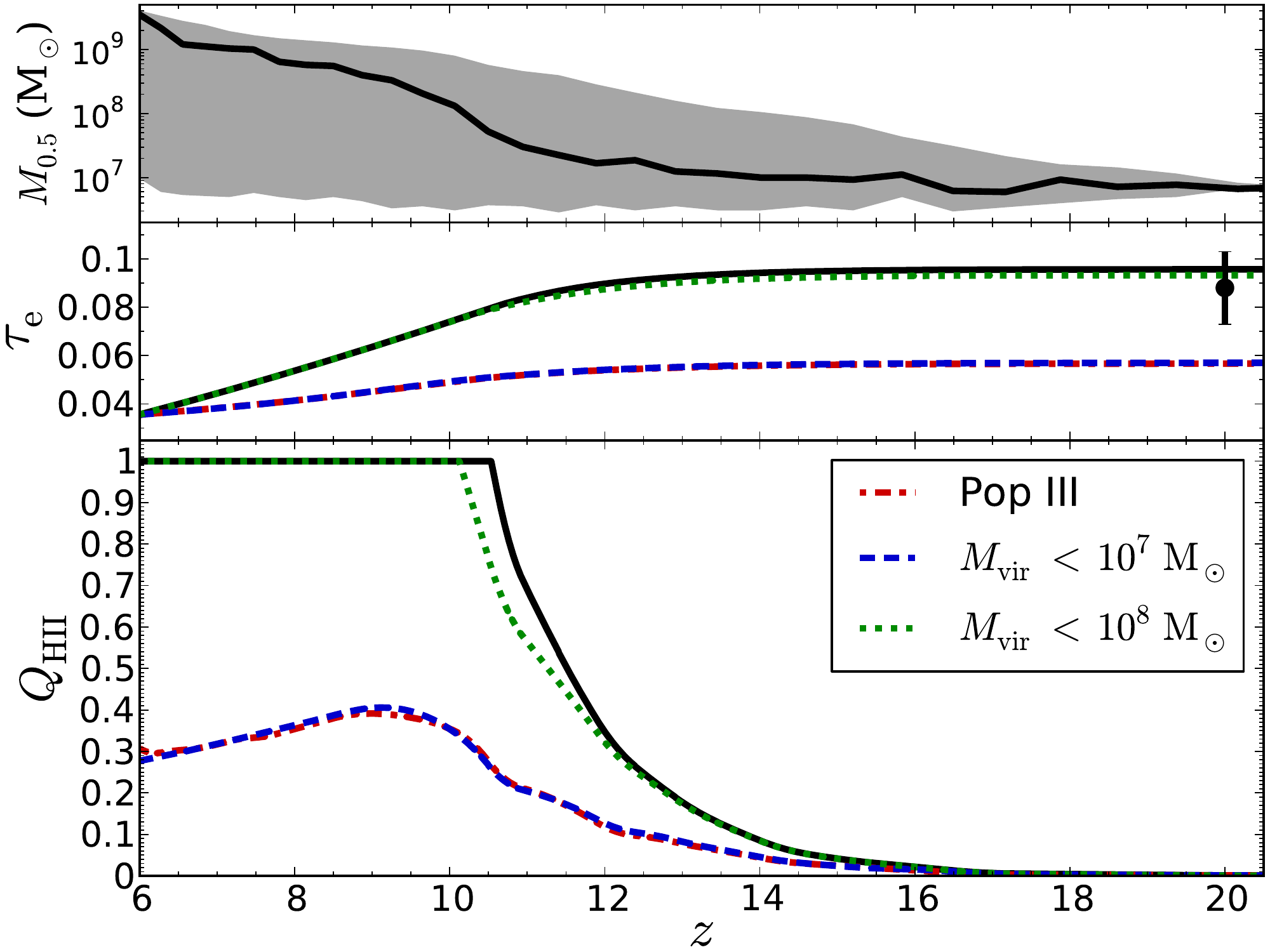}
  \caption{The contribution of proto-galaxies to reionisation. In all panels the black solid lines represent all haloes, while the blue dashed and green dotted lines show the contribution of haloes with masses below $10^7 \, \mathrm{M}_{\odot}$ and $10^8 \, \mathrm{M}_{\odot}$, respectively. The red dash-dot lines represent the contribution of Pop III stars. {\it Top panel:} The halo mass below which 50\% of the ionising photons is produced, $M_{0.5}$, at each redshift. The grey area represents the range of halo masses in which star formation is taking place. {\it Middle panel:} The optical depth for Thomson scattering, $\tau_{\mathrm{e}}$ as function of redshift. The data point represents the value of $\tau_{\mathrm{e}}$ as found by measurements with the WMAP satellite \citep{2011ApJS..192...18K}. {\it Bottom panel:} The volume filling fraction of ionised hydrogen, $Q_{\mathrm{H\,II}}$, as function of redshift. }
  \label{fig_Q}
\end{figure}

Haloes with masses below $10^8 \, \mathrm{M}_{\odot}$ dominate the ionising photon budget at redshifts higher than 10, due to their high escape fractions and high abundance. After redshift 10, the ionising emissivity of the proto-galaxies in these haloes drops as a result of suppression of star formation by the uniform UV-background. The background heating only suppresses star formation in haloes that do not contain enough dense gas to shield against the radiation. This counteracts the effect of the high escape fraction, since ionising radiation is mainly produced by massive, young stars and suppression of star formation results in little or no ionising radiation being produced (see Fig. \ref{fig_Q}). Low-mass haloes therefore only contribute to the ionising photon budget when the part of the Universe they live in is still neutral, while more massive haloes have enough dense gas to shield against the external radiation and continue to form stars. 

In Fig. \ref{fig_Q} we show the evolution of $Q_\mathrm{H \, II}$ and $\tau_\mathrm{e}$ with redshift. Reionisation is complete at redshift 10.5, with a duration (defined as the redshift interval in which $Q_\mathrm{H \, II}$ changes from 20\% to 80 \%) of $\Delta z \approx 2.1$. This is before the lower limit of the end of reionisation set by the measurements of quasar spectra. The dashed curve shows the reionisation history including only proto-galaxies residing in $< 10^7 \, \mathrm{M}_{\odot}$ haloes, which do not produce enough photons to reionise the Universe. The bulk of photons is produced by proto-galaxies in haloes with masses between $10^7 \, \mathrm{M}_{\odot}$ and $10^8 \, \mathrm{M}_{\odot}$. Reionisation is only delayed by $\Delta z \approx 0.5$ if we exclude all sources in haloes with masses larger than $10^8 \, \mathrm{M}_{\odot}$. 

The integrated Thomson scattering optical depth that we find in our model is $\tau_\mathrm{e} = 0.096$, which is well within the error bars from the WMAP measurement  \citep{2011ApJS..192...18K}. This shows that the reionisation history we find is consistent with the two main observational constraints of reionisation, the absorption features in high-redshift quasars and the Thomson optical depth as observed by WMAP.

The contribution of Pop III stars to reionisation is negligible in our simulations. Although these sources produce copious amounts of ionising photons \citep{Schaerer:2002bm}, they are short-lived and not abundant enough to contribute significantly to reionisation. The contribution of Pop III stars to the total photon budget is exceeded by metal-enriched Pop II stars at redshifts below 15. We thus conclude that although reionisation started with the appearance of the first Pop III stars, they did not contribute significantly to reionisation on global scale. 

In the top panel in Fig. \ref{fig_Q} we show the halo mass below which 50\% of the ionising photons are produced as function of redshift. At redshifts higher than $10$, half of the ionising photons are produced by proto-galaxies in haloes with masses between $10^7 \, \mathrm{M}_{\odot}$ and $10^8 \, \mathrm{M}_{\odot}$. After this redshift, haloes with higher masses take over the photon production, because star formation is suppressed in the lower mass haloes due to the UV-background.

\begin{figure} %the cumulative number of photons produced per baryon in the Universe as function of halo mass
  \includegraphics[width=\columnwidth]{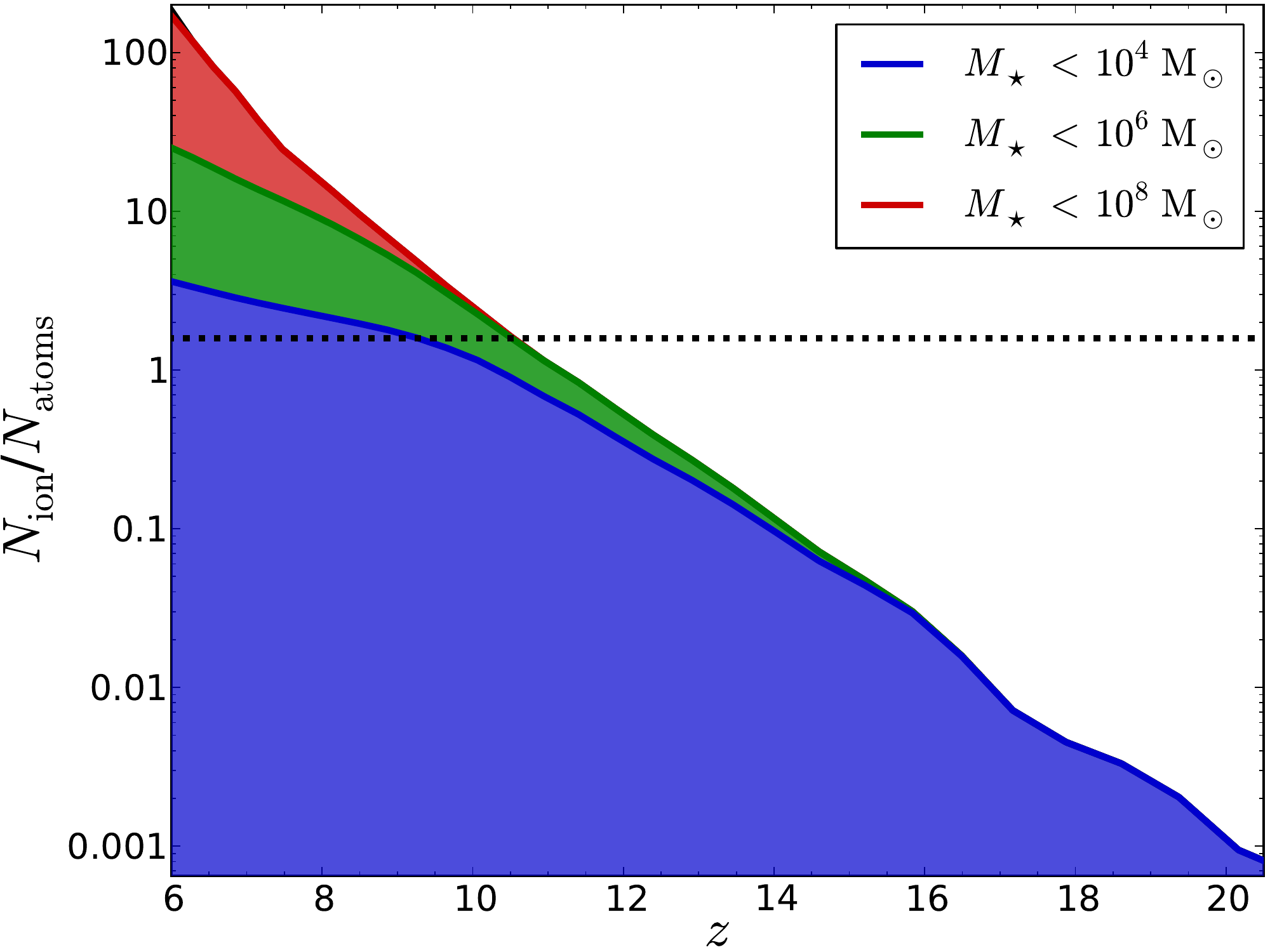}
  \caption{The cumulative number of ionising photons per baryon as function of redshift. Different colors denote the contribution of proto-galaxies below a certain stellar mass. In our computation full reionisation requires $1.6$ photons per baryon, which is represented by the black dotted line.} 
  \label{fig_NIonBaryon}
\end{figure}

To get a better picture of the mass of galaxies mostly contributing to reionisation, we show in Fig. \ref{fig_NIonBaryon} the cumulative number of photons per baryon produced by proto-galaxies below a certain stellar mass as function of redshift. Given our estimates of the ionising emissivity and the clumping factor, reionisation requires approximately 1.6 photons per baryon. Proto-galaxies with stellar masses below $10^6 \, \mathrm{M}_{\odot}$ have produced this number of photons by redshift 10, with most ionising photons being produced by proto-galaxies with stellar masses between $10^5 \, \mathrm{M}_{\odot}$ and $10^6 \, \mathrm{M}_{\odot}$.

\section{The observability of the sources of reionisation}

\begin{figure} %the observability of proto-galaxies in continuum and three recombination lines
  \includegraphics[width=\columnwidth]{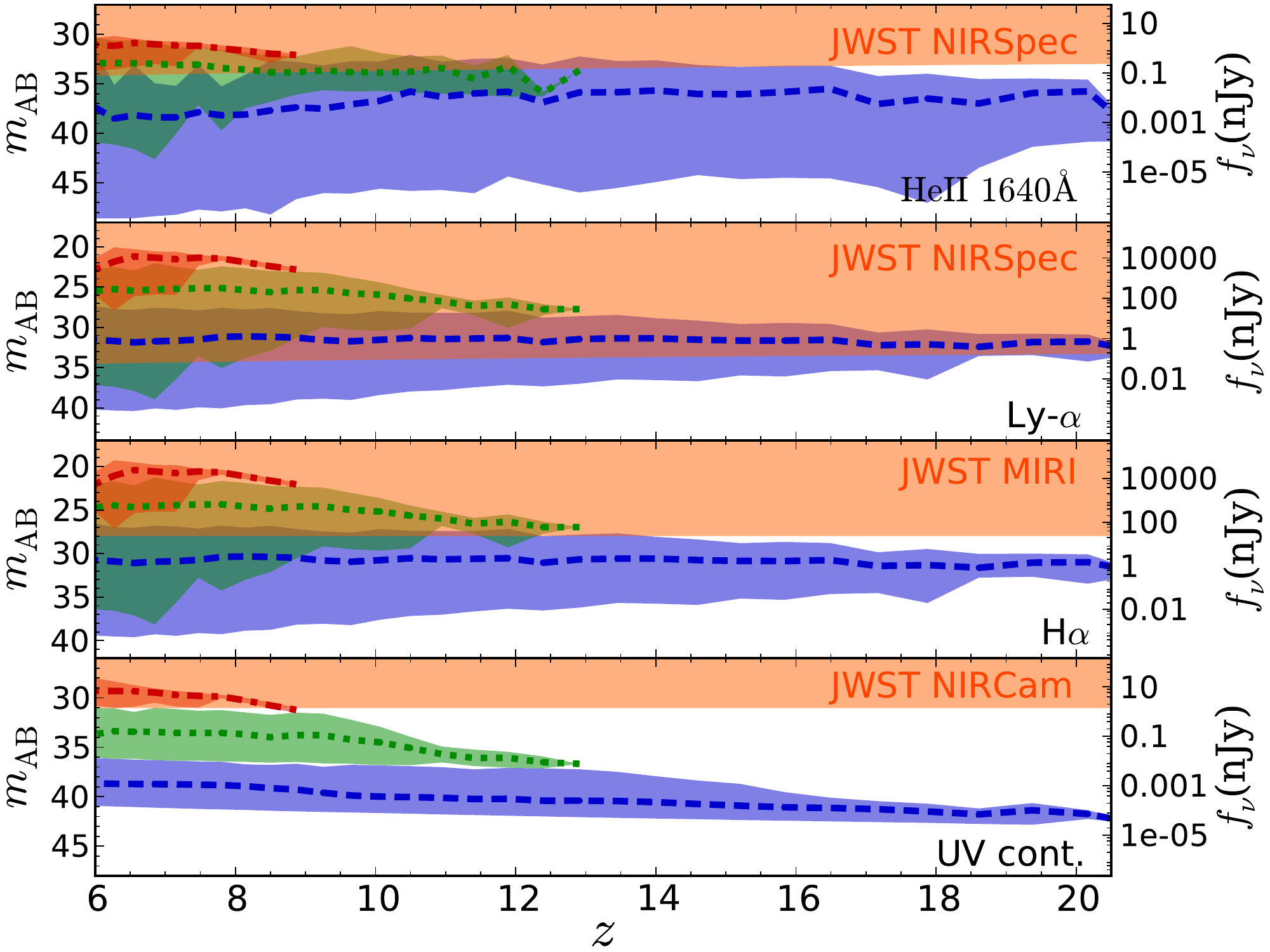}
  \caption{The observability of the sources of reionisation in the UV continuum and three recombination lines. In all plots colors are similar to the colors in Fig. \ref{fig_NIonBaryon}. The shaded area denotes the maximum and minimum brightness and the orange area covers the range detectible by the relevant instrument on JWST.
  } 
  \label{fig_obs}
\end{figure}

One of the science goals of the {\it James Webb Space Telescope} (JWST) is to study the galaxy population during reionisation \citep{2006SSRv..123..485G}. The deepest NIRCam survey will search for high-redshift objects with exposure times of $10^6$ seconds, resulting in a flux limit of $1.4 \, \mathrm{nJy}$ at $2 \, \mu\mathrm{m}$. This can be converted into AB magnitude using
\begin{equation}
  m_{\mathrm{AB}} = 31.4 - 2.5\log(f_{\nu}),
\end{equation}  
resulting in a magnitude limit of $m_{\mathrm{AB}} = 31.0$. In addition, JWST will search for recombination line radiation from the first galaxies using the MIRI and NIRspec instruments. At $5.6 \, \mu \mathrm{m}$, the limiting flux of the MIRI instrument is $23$nJy or $m_{\mathrm{AB}} = 28$, while NIRSpec will measure line intensities down to $2 \times 10^{-19} \, \mathrm{erg} \, \mathrm{cm}^{-2} \, \mathrm{s}^{-1}$ with a resolution $R=1000$.

We determine the flux that the NIRCam survey would receive from the proto-galaxies in our simulation by computing the spectra of the Pop II and Pop III stars for radiation with energy below the ionisation energy of hydrogen. In Fig. \ref{fig_obs} we show the redshifted flux from the sources in the  $2 \, \mu\mathrm{m}$ band, although due to the flatness of the spectrum in this wavelength range the results do not depend sensitively on the choice of wavelength. Reionisation is driven by proto-galaxies with stellar masses around $10^6 \, \mathrm{M}_{\odot}$. At redshift 10.5, when the proto-galaxies have produced enough photons to reionise the volume, these sources are not brighter than $f_{\nu} = 0.09 \,\,\mathrm{nJy}$ or $m_{\mathrm{AB}} = 34$ , making it impossible for JWST to observe them. If we rescale the flux limits according to $f_{\mathrm{lim}} \propto 1/\sqrt{t_{\mathrm{exp}}}$, exposure times of $2 \times 10^8 \mathrm{s}$ are necessary to observe these sources. Since we do not account for dust attenuation between the proto-galaxy and the observer, we can only give upper limits of the flux.

In addition, we show in Fig. \ref{fig_obs} the brightness of the proto-galaxies in our simulation in three emission lines, $\mathrm{H}\alpha$, Ly-$\alpha$ and $\mathrm{He II} \, 1640 \mbox{\AA}$. The flux in these lines is given by \citep{Johnson:2009gb}
\begin{equation}
  f(\lambda_{\mathrm{obs}}) = \frac{\ell_{\mathrm{em}} \lambda_{\mathrm{em}} (1+z) R}{4 \pi c D_{\mathrm{L}}^2(z)},
\end{equation}
where $\ell_{\mathrm{em}}$ is the luminosity along the line of sight, $D_{\mathrm{L}}^2(z)$ is the luminosity distance at redshift $z$, $c$ is the light speed and $R=\lambda/\Delta \lambda$ is the spectral resolution. Assuming that the galaxy is unresolved, we compute $\ell_{\mathrm{em}}$ for every proto-galaxy by summing up the contribution to the emissivity from every gas particle.

JWST will observe the $\mathrm{H}\alpha$ line with the MIRI instrument. At redshift $12$ and below, proto-galaxies of $M_{\star} \ge 10^6 \, \mathrm{M}_{\odot}$ are bright enough to be detected with MIRI. The Ly-$\alpha$ and $\mathrm{He II} \, 1640 \mbox{\AA}$  lines will be observed with the NIRSpec instrument. The greater sensitivity of the NIRSpec instrument could in principle make it possible to observe the Ly-$\alpha$ emission line from all sources with $M_{\star} \ge 10^4 \, \mathrm{M}_{\odot}$ at redshift 20 and below. However, this line is difficult to detect at these high redshifts due to the large Gunn-Peterson optical depth. Scattering off interstellar neutral gas could make the line observable even through a fully neutral IGM \citep{2010MNRAS.408..352D}. The flux in the $\mathrm{He II} \, 1640 \mbox{\AA}$ line is always lower than the $\mathrm{H}\alpha$ flux. It is therefore most likely that the sources of reionisation will be observed in the $\mathrm{H}\alpha$ or Ly-$\alpha$ line.

\section{Conclusions and discussion}

We have presented high-resolution cosmological simulations of galaxy formation, from which we calculated the ionising photon production and escape fraction of a large statistical sample of galaxies. From these simulations we computed the contribution of proto-galaxies to cosmic reionisation in a self-consistent way. Our main findings are:
\begin{itemize}

\item Reionisation is primarily driven by proto-galaxies in dark-matter haloes with masses between $10^7 \, \mathrm{M}_{\odot}$ and $10^8 \, \mathrm{M}_{\odot}$, which have very high escape fractions, because supernova feedback efficiently acts to clear away gas from the sites of star formation. 

\item Star formation in these haloes is suppressed by UV-feedback, these proto-galaxies therefore only contribute to reionisation when the part of the Universe they live in is still neutral.

\item After reionisation the Universe is kept ionised by massive galaxies with lower escape fractions, that start appearing more frequently. 

\item We find a strong mass and redshift dependence of the escape fraction that suggests that galaxies above the observational limits of present surveys do not contribute enough photons to drive reionisation.

\item Pop III stars do not contribute significantly to the reionisation process.

\item There is great prospect that JWST will observe the proto-galaxies that reionised the Universe in the $\mathrm{H}\alpha$ and Ly-$\alpha$ recombination lines.

\end{itemize}
In our galaxy formation simulation radiative feedback on the halo gas from sources within the halo is neglected. Radiative feedback is capable of evacuating gas from haloes with $M_{\mathrm{vir}} \lesssim 10^7 \, \mathrm{M}_{\odot}$ \citep{Wise:2009fn}, thereby suppressing further star formation in the halo. We find that supernova feedback has the same effect, albeit with a short delay. Since the time it takes for the dense, cold gas in the halo to be converted into stars ($\sim 1 \, \mathrm{Gyr}$) is much larger than the lifetime of the massive stars that end their lives as supernova ($\sim 1 \, \mathrm{Myr}$ for pair-instability supernovae), this delay will not affect our results significantly.

We model reionisation with a uniform UV-background, disregarding the contribution from local sources. We could therefore be underestimating the suppression of star formation in haloes that are not able to shield against the ionising radiation. This likely occurs in haloes with masses $M_{\mathrm{vir}} \lesssim 10^9 \, \mathrm{M}_{\odot}$ \citep{Okamoto:2008ha,Hasegawa:2012wu}. Since this will only affect proto-galaxies in a region of the Universe that has already been ionised, this will not change our conclusions. However, it could delay the completion of reionisation due to suppression of clustered low-mass sources and reduce the number of photons available for reionisation. Assuming all star formation in haloes with $M_{\mathrm{vir}} < 10^9 \, \mathrm{M}_{\odot}$ is suppressed after reionisation brings the cumulative number of ionising photons per baryon at redshift 6 in our simulation close to the observed value \citep{Bolton:2007gc}. 

The resolution in our simulation is around 6 physical pc at redshift 15, which is high enough to resolve regular and giant molecular clouds at the sites of star formation. However,  we do not resolve the birth cloud of the stellar population, which would lower the escape fraction. Since this equally applies to all haloes in the volume, it does not affect our conclusion on the mass range that contributes most photons to reionisation. The effect of a lower escape fraction can be estimated by scaling the curves in Fig. \ref{fig_NIonBaryon} accordingly.

The volume of our simulation is 4 Mpc, which could possibly bias our results. The initial conditions for the cosmological simulation were chosen to avoid biased regions with high $\sigma$ peaks. This is reflected in the mass function of dark matter haloes in our volume, which is in agreement with the average of the Sheth-Tormen mass function \citep{Sheth:2002fn}. Although we cannot accurately sample the high-mass end of the halo mass function, the strong mass-dependence we find for the escape fraction indicates that the contribution from these rare high-mass sources is small. 

Our results show that reionisation is initially a local process driven by the appearance of many low-mass haloes which do not cluster strongly and ionise their neighbourhood until \Hii regions overlap and cover larger cosmic volumes.

\section*{Acknowledgments}

We would like to thank Jarrett Johnson and the TMoX group for helpful discussions and Jim Dunlop and Chael Kruip for comments on an earlier draft. C.D.V. acknowledges support by Marie Curie Reintegration Grant FP7-RG-256573.

\bibliographystyle{mn2e} % style aa.bst
%\bibliography{../allPapers} % your references Yourfile.bib

\bsp

\label{lastpage}

\end{document}